# Superconducting parallel nanowire detector with photon number resolving functionality

F. Marsili[1,4,*], D. Bitauld[1,4], A. Fiore[1,4], A. Gaggero[2], R. Leoni[2], F. Mattioli[2], A.Divochiy[3], A.Korneev[3], V.Seleznev[3], N.Kaurova[3], O.Minaeva[3], G.Goltsman[3]

[1]Ecole Polytechnique Fédérale de Lausanne (EPFL), Institute of Photonics and Quantum Electronics (IPEQ), Station 3, CH-1015 Lausanne, Switzerland;
[2]Istituto di Fotonica e Nanotecnologie (IFN), CNR, via Cineto Romano 42, 00156 Roma, Italy;
[3]Moscow State Pedagogical University (MSPU), Department of Physics, 119992 Moscow, Russian Federation ;
[4]Present address: COBRA Research Institute, Eindhoven University of Technology, P.O. Box 513, NL-5600MB Eindhoven, The Netherlands

We present a new photon number resolving detector (PNR), the Parallel Nanowire Detector (PND), which uses spatial multiplexing on a subwavelength scale to provide a single electrical output proportional to the photon number. The basic structure of the PND is the parallel connection of several NbN superconducting nanowires (≈100 nm-wide, few nm-thick), folded in a meander pattern. Electrical and optical equivalents of the device were developed in order to gain insight on its working principle. PNDs were fabricated on 3-4 nm thick NbN films grown on sapphire (substrate temperature $T_S$=900°C) or MgO ($T_S$=400°C) substrates by reactive magnetron sputtering in an $Ar/N_2$ gas mixture. The device performance was characterized in terms of speed and sensitivity. The photoresponse shows a full width at half maximum (FWHM) as low as 660ps. PNDs showed counting performance at 80 MHz repetition rate. Building the histograms of the photoresponse peak, no multiplication noise buildup is observable and a one photon quantum efficiency can be estimated to be η~3% (at 700 nm wavelength and 4.2 K temperature). The PND significantly outperforms existing PNR detectors in terms of simplicity, sensitivity, speed, and multiplication noise.

---

*Corresponding author. Email: <u>francesco.marsili@epfl.ch</u>

# 1. Introduction

Optical detection involves converting an optical signal into an electrical signal. Most optical detectors are operated in a linear mode, where the current or voltage output is proportional to the optical power, i.e. to the incident photon flux. The main limitation in the sensitivity (minimum measurable photon flux) of these linear detectors is the ability to extract a small photocurrent from amplifier noise in a given bandwidth. For example, avalanche photodiodes commonly used in fiber-based optical communications have sensitivities of few hundreds of photons in a 100 ps detection window. When higher sensitivities are needed, single-photon detectors are often used, which operate in a strongly nonlinear mode. The typical example is the avalanche photodiode biased in the Geiger mode, where a single photon is sufficient to generate an avalanche breakdown in a semiconductor junction, which produces a large current pulse. In APDs, as in most single-photon detectors, a pulse containing more than one photon produces the same voltage pulse as a single-photon pulse, which implies that it is not possible to directly measure the number n of photons in a pulse, if the pulse duration is smaller than the detector response time. However there is a common need for photon-number resolution, particularly in quantum optics, where single-photon and entangled two- and n-photon states are routinely produced. The typical solution consists of splitting the pulse in several paths and correlating the output of several single-photon detectors, one on each path – for example a 2-photon pulse produces two clicks with 50% probability (and a false measurement in the remaining cases). The historical prototype of these correlation techniques is the Hanbury-Brown and Twiss experiment [1]. However, this approach becomes impractical if states with n>2 must be accurately measured, as the number of detectors should be >>n. Among the approaches proposed so far to PNR detection, detectors based on charge-integration or field-effect transistors [2-4] are affected by long integration times, leading to bandwidths <1 MHz. Transition edge sensors (TES [5]) are able of directly measuring the total energy, hence the number of photons, in a pulse, but they operate at 100 mK and show long response times (the shortest record reported is several hundreds of nanoseconds [6]). Approaches based on photomultipliers (PMTs) [7] and avalanche diodes (APDs), such as the visible light photon counter (VLPC) [8, 9], 2D arrays of APDs [10, 11] and time-multiplexed detectors [12, 13] are not sensitive or are plagued by high dark count rate and long dead times in the telecommunication spectral windows. Arrays of SPDs additionally involve complex read-out schemes [11] or separate contacts, amplification and discrimination [14]. We recently demonstrated an alternative approach, the Parallel Nanowire Detector (PND), which uses spatial multiplexing on a subwavelength scale to provide a single electrical output proportional to the photon number. The device presented significantly outperforms existing PNR detectors in terms of simplicity, sensitivity, speed, and multiplication noise. A short description of the experimental results was reported in [15]. Here we present some new experimental results and their more detailed analytical discussion.

# 2. Photon Number Resolution principle

The basic structure of the PND is the parallel connection of $N$ superconducting nanowires ($N$-PND). The detecting element is a few nm-thick, ≈100 nm-wide NbN wire folded in a meander pattern (Fig. 1a). Each branch acts as a superconducting single photon detector (SSPD) [16]. If a superconducting nanowire is biased close to its critical current, the absorption of a photon causes the formation of a normal barrier across its cross section and the bias current is pushed to the external circuit. In the parallel configuration proposed here, the currents from different wires can sum up on the external load, producing an output voltage pulse proportional to the number of photons.

When the device in the superconducting state is biased using a voltage source in series with a bias resistor, the current $I_B$ will spread equally between the $N$ branches, so $I^i = I_B / N$ for $i=1,..,N$. If all the $N$ nanowires are exactly the same, all with a critical current $I_C$, the superconducting to normal transition of the whole structure takes place when the bias current through the device exceeds $I_C^{tot} = N \cdot I_C$. As the nanowires are differently constricted, the bias current can be increased only till each current flowing through one of the N nanowires exceeds the critical value for that particular nanowire $I_C^i$, so $I_C^{tot} = \sum_{i=1}^{N} I_C^i$. Let the $k$-th nanowire be the most constricted and its $I_C^k$ be the lowest. While increasing $I_B$, as long as $I^i < I_C^i$, $I^i = I_B / N$ for all the branches, but as the current through the $k$-th nanowire approaches the critical value $I^k = I_B/N \sim I_C^k$, it is fixed to that value. If $I^k$ is further increased beyond $I_C^k$, the $k$-th nanowire switches to the dissipative hot spot state. A voltage drop then appears across it and all the other still superconducting branches, which drains enough current out of the $k$-th nanowire to drive it back to the superconducting state. The same argument also applies to the other nanowires, so in this parallel structure it is possible to bias each branch very close to its own value of $I_C$. Let $\eta_i = \eta(i^{(i)})$ (with $i^{(i)} = I^i / I_C^i$) be the current-depending quantum efficiencies of a nanowire [16] (the nanowires have a different critical currents, being differently constricted [17]). When the device is biased very close to $I_C^{tot}$, each nanowire will then give its best performance, its $\eta_i$ being maximum as $i^{(i)} \sim 1$.

When a photon reaches the $i$-th nanowire biased at a current $I^i$, it will cause the creation of a normal barrier across the entire cross section of the nanowire with a probability $\eta_i$. Because of the sudden increase in the resistance of the nanowire, its current is then redistributed between the other $N-1$ nanowires in parallel and the 50 Ω input resistance $R_A$ of the high frequency amplifier. The peak value of the output current pulse flowing through $R_A$ is then $I_{out}(1) \sim [I_B/N - (N-1) \cdot \delta I_{lk}]$, where $\delta I_{lk}$ is the leakage current drained by each of the still superconducting nanowires. This argument yields that if $m$ ($m \leq N$) photons are simultaneously (in a time interval much shorter than the current relaxation time) absorbed in $m$ distinct nanowires, the currents from different branches can sum up on the external load producing a current pulse of height $I_{out}(m) \sim [m/N \cdot I_B - m(N-m) \cdot \delta I_{lk}]$.

The device shows photon-number-resolving (PNR) capability if the remaining wires do not shunt the load, i.e. if $\delta I_{lk} << I_B / N$, so that $I_{out}(m) \sim m\ I_{out}(1)$. The leakage current is also undesirable because it lowers the signal available for amplification and temporary increases the $I^i$'s, eventually driving other nanowires normal (if $(I^i + m \cdot \delta I_{lk}) > I_C^i$, where $m$ is the number of detected photons). Consequently $\delta I_{lk}$ limits the maximum bias current allowed for the stable operation of the device and then the quantum efficiencies of the branches. The leakage current depends on the ratio between the impedance of a section $Z_B$ and $R_A$. As $Z_B$ is due just to the kinetic inductance of the superconducting nanowire $L_{kin}$ [18], the trade off between $\delta I_{lk}$ and the speed of the device is very narrow. In order to relax the design constraints, a PND with bias resistors ($R_0$) integrated in series to the nanowires in each section (PND-R) was developed (fig. 1b) [15]. In a PND-R $Z_B = j\omega L_{kin} + R_0$, so carefully engineering the geometrical dimensions of the nanowire and of the bias resistor, $\delta I_{lk}$ can be minimized keeping the time response of the device below the nanosecond ($L_{kin}$ in the 100 nH range) and the bias condition stable ($R_0$ in the 50 Ω range).

## 3. Modelling

### 3.1. Electrical model

The time evolution of the device after photon absorption can be simulated using the equivalent circuit of fig. 2a. Each nanowire is modeled as the series of an inductance $L_{kin}$ accounting for its kinetic inductance, a switch $S_l$ which opens on the hotspot resistance $R_{hs}$, simulating the absorption of a photon, and a switch $S_c$ which opens on the resistance $R_n$ which represents the resistance of a thermal domain that appears as the current flowing through the wire exceeds the critical value. Each section is the series connection of a nanowire and of a bias resistor $R_0$. The device is connected through a bias T to the bias voltage source $V_B$ and to the 50 Ω matched transmission line, terminating on the radio-frequency amplifier with input resistance $R_A$=50 Ω.

Figure 2b shows the simulation result for the time evolution of the currents flowing through the sections of a PND with 4 sections and integrated resistors (4-PND-R) and through $R_A$, under illumination with pulses containing 1-4 photons. We assume here an ideal situation where the photons in each pulse are never absorbed on the same nanowire and the quantum efficiencies are unitary, so the number of incident photons is equal to the number of sections making a transition. The sections absorbing the photon experience a large drop in their current, the others experience an increment which is a multiple of $\delta I_{lk}$.

### 3.2. Analytical model

The photon number probability distribution $Q(n)$ measured with a PNR detector is related to the incoming distribution $S(m)$ by the relation:

$$Q(n) = \sum_m P(n|m) \cdot S(m) \qquad (1)$$

where $P(n|m)$ is the probability that $n$ photons are detected when $m$ are sent to the device. Let $P^N = \left[ P^N_{nm} \right]$, with $P^N_{nm} = P^N(n|m)$, be the matrix of the conditional probabilities for an $N$-PND. Assuming that the illumination of the device is uniform, the parallel connection of $N$ nanowires can be considered equivalent to a balanced lossless $N$ port beam splitter, every channel terminating with a single photon detector (SPD) (fig. 3a). Each incoming photon is then equally likely to reach one of the $N$ SPDs (with a probability 1/$N$). Each SPD can detect a photon with a probability $\eta_i$ ($i=1,..,N$) different from all the others, and gives the same response for any number ($m \geq 1$) of photons detected (fig. 3b). The number of SPDs firing then gives the measured photon number. Following[13], two classes of terms in $P^N$ can be calculated directly, the others being derived from these by a recursion relation. These terms are the probabilities $P^N_{m,m}$ that all the $m \leq N$ photons sent are detected and $P^N_{0,m}$ that no photons are detected when $m$ are sent.

In the case of zero detections, $P^N_{0,m}$ is given by:

$$P^N_{0,m} = \sum_{i_1=1,\ldots,i_m=1}^{N} \left[ \frac{1-\eta_{i_1}}{N} \cdot \ldots \cdot \frac{1-\eta_{i_m}}{N} \right] \qquad (2)$$

which assumes that a photon incident in the $i$-th nanowire fails to be detected with an independent probability of $(1-\eta_i)$. The sum in (2) accounts for all the possible combinations when taking $m$ elements in an ensemble of $N$ with order and with repetition ( permutations with repetitions). This because more than one photon can hit the same stripe (which gives the repetition), and the photons are considered distinguishable (which gives the order). The sum has then $N^m$ terms. If all the stripes are the same ($\eta_i=\eta$), (2) reduces to (A1) in[13].

In the case that all the photons are detected, since $m$ photons must reach $m$ distinct nanowires:

$$P_{m,m}^N = \sum_{\substack{i_1=1,\ldots,i_m=1 \\ i_p \neq i_q \text{ for } p \neq q}}^{N} \left[ \frac{\eta_{i_1}}{N} \cdot \ldots \cdot \frac{\eta_{i_m}}{N} \right] \quad \text{for } m \leq N \qquad (3)$$

The sum in (3) accounts for all the possible combinations when taking $m$ elements in an ensemble of $N$ with order and without repetition (permutations without repetitions). This because only one photon can hit the same stripe (which gives the non-repetition), and the photons are considered distinguishable (which gives the order). The sum has then $N!/(N-m)!$ terms. If all the stripes are the same ($\eta_i=\eta$) (3) reduces to (A2) in[13].

The recursion relation for $P_{nm}^N$ is:

$$P_{n,m}^N = P_{n,m-1}^N \left[ \frac{n}{N} + \frac{n!}{N!} \cdot \sum_{\substack{i_1=1,\ldots,i_{N-n}=1 \\ i_p \neq i_q \text{ for } p \neq q}}^{N} \left( \frac{1-\eta_{i_1}}{N} + \ldots + \frac{1-\eta_{i_{N-n}}}{N} \right) \right] + $$
$$ + P_{n-1,m-1}^N \left[ \frac{(n-1)!}{N!} \cdot \sum_{\substack{i_1=1,\ldots,i_{N-(n-1)}=1 \\ i_p \neq i_q \text{ for } p \neq q}}^{N} \left( \frac{1-\eta_{i_1}}{N} + \ldots + \frac{1-\eta_{i_{N-(n-1)}}}{N} \right) \right] \qquad (4)$$

The first term on the right-hand side of (4) is the probability that $n$ photons are detected when $m$-1 are sent, times the probability that the $m^{th}$ photon reaches one of the $n$ nanowires already occupied (first term in the square brackets) or that it fails to be detected reaching one of the $N$-$n$ unoccupied nanowires (second term in the square brackets). To clarify how the latter probability is derived, it is sufficient to consider a particular configuration $k$ (see fig. 3b) of $n$ *fired* stripes (which have already detected a photon) and $N$-$n$ *unfired* stripes (still active). The probability that the incoming photon will not be detected when incident on any of the $N$-$n$ *unfired* stripes is then written as:

$$p_k = \frac{1-\eta_{i_1}}{N} + \ldots + \frac{1-\eta_{i_{N-n}}}{N} \qquad (5)$$

where $i_1 \ldots i_{N-n}$ are the $N$-$n$ stripes active in the $k^{th}$ configuration of *(n)fired-(N-n)unfired* stripes considered. So a mean must be calculated on all the possible *(n)fired-(N-n)unfired* configurations for the $N$ stripes. Let $H$ be the number of all these configurations. The mean is then calculated summing $H$ terms of the type (5), and dividing by $H$: $1/H \cdot \sum_{k=1}^{H} p_k$. $H$ is the number of permutations without repetitions of $N$-$n$ elements in an ensemble of $N$, and it is given by the binomial coefficient: $H = N!/(N-(N-n))! = N!/n!$.

The second term on the right-hand side of (4) is the probability that $n$-1 photons are detected when $m$-1 are sent times the probability that the $m^{th}$ photon reaches one of the $N$-($n$-1) unoccupied nanowires and it is detected. In the limit $\eta_i = \eta$ for $i$=1,...,$N$, the recursion relation agrees with that given in [13].

### 3.3. Monte Carlo simulation

In order to prove the consistency of the analytical model, the detected photon number probability distribution $Q(n)$ calculated from $P^N$ by (1) was cross-checked with the $Q(n)$ resulting from a Monte Carlo simulation. The input parameters of the simulation are the incoming photon number probability distribution $S(m)$, the number of parallel stripes $N$, and the vector of the quantum efficiencies $\bar{\eta}$. The detected photon number distribution was calculated using the following algorithm:

1. All the $N$ sections of the $N$-PND are marked as *unfired*. The number of photons incident on the detector $n$ is determined measuring a random variable with probability distribution $S(m)$.
2. Each photon hits the $i^{th}$ section with uniform probability ($1/N$). If this section has not already detected a photon, it will be marked as *fired* with a probability $\eta_i$.
3. The number $n$ of *fired* sections is kept as the result of this iteration for statistical analysis.
4. Repeat step 1. to 3. for $10^6$ times.

### 4. Fabrication

NbN films 3-4 nm thick were grown on sapphire (substrate temperature $T_S$=900°C [19, 20]) or MgO ($T_S$=400°C [21]) substrates by reactive magnetron sputtering in an argon–nitrogen gas mixture. Using an optimized sputtering technique, our NbN samples exhibited a superconducting transition temperature of $Tc$ =10.5 K for 40-Å-thick films. The superconducting transition width was equal to $\Delta Tc$ = 0.3 K.

Both the designs with and without the integrated bias resistors were implemented. Scanning-electron microscope (SEM) pictures of a 14-PND and an 8-PND-R fabricated on MgO are shown in fig. 1a and 1b, respectively. Detector size ranges from 5x5 μm² to 10x10 μm² with the number of parallel branches varying from 4 to 14. The nanowires are 100 to 120 nm wide and the fill factor of the meander is 40 to 60%. The length of each nanowire ranges from 25 to 100 μm.

For the devices on MgO, the three nanolithography steps needed to fabricate the structure have been carried out by using an electron beam lithography (EBL) system equipped with a field emission gun (acceleration voltage 100 kV, 20 nm resolution). In the first step pads (patterned as a 50 Ω coplanar transmission line) and alignment markers are fabricated by lift off of a Ti-Au film (60 nm Au on 10nm Ti, deposited by e-gun evaporation) through a Polymethyl Methacrylate (PMMA, a positive tone electronic resist) stencil mask. In the second step, an hydrogen silsesquioxane (HSQ FOX-14, a negative tone electronic resist) mask properly aligned to the previous layer is defined reproducing the meander pattern. All the unwanted material, i.e. the material not covered by the HSQ mask and the Ti/Au film, is removed by using a fluorine based reactive ion etching (RIE). Finally, with the third step the bias resistors (85nm AuPd alloy, 50%-each in weight) aligned with the two previous layers are fabricated by lift off via a PMMA stencil mask. The successive alignments of the two last layers were performed with an error of the order of 100 nm.

## 5. Measurement setup

Electrical and optical characterizations have been performed in a cryogenic probe station with an optical window and in cryogenic dipsticks. Bias current was supplied through the DC port of a 10MHz-4GHz bandwidth bias-T connected to a low noise voltage source in series with a bias resistor. The AC port of the bias-T was connected to the room-temperature, low-noise amplifiers. The amplified signal was fed either to a 1 GHz bandwidth single shot oscilloscope, a 40 GHz bandwidth sampling oscilloscope, or a 150MHz bandwidth counter for time resolved measurements and statistical analysis. The devices were optically tested using a fiber-pigtailed, gain-switched laser diode at 1.3 μm wavelength (100ps-long pulses, repetition rate 26 MHz), a mode-lock Ti:sapphire laser at 700 nm wavelength (40ps-long pulses, repetition rate 80 MHz), or an 850 nm GaAs pulsed laser (30 ps-long pulses, repetition rate 100 kHz).

Throughout the paper, the efficiency η is defined with respect to the photon flux incident on the device area, typically 10x10 μm$^2$.

In the cryogenic probe station (Janis) the devices were tested at a temperature T=5 K. Electrical contact was realized by a cooled 50 Ω microwave probe attached to a micromanipulator, and connected by a coaxial line to the room-temperature circuitry. The light was fed to the PNDs through a single-mode optical fiber coupled with a long working distance objective, allowing the illumination of a single detector.

In the cryogenic dipsticks the devices were tested at 4.2 K or 2 K. The light was sent through a single-mode optical fiber either put in direct contact and carefully aligned with the active area of a single device or coupled with a short focal length lens, placed far from the plane of the chip in order to ensure uniform illumination. The number of incident photons per device area was estimated with an error of 5 %.

## 6. Device Characterization

### 6.1. Speed performance

The photoresponse of a 10x10 μm$^2$ 4-PND-R probed with light at 1.3 μm was recorded by the 40 GHz sampling oscilloscope (fig. 4a). All four possible amplitudes can be observed. The pulses show a full width at half maximum (FWHM) as low as 660ps. PNDs showed counting performance when probed with light at 80 MHz repetition rate (fig. 4b), outperforming any existing PNR detector at telecom wavelength by three orders of magnitude. Indeed, PNDs have by design a reduced recovery time even compared to traditional SSPDs [19]. After the relaxation of the hot spot, the current through the nanowire recovers with the characteristic time $\tau=L_{eq}/R_A$, where $L_{eq}$ is the equivalent inductance seen by $R_A$. For an SSPD, the equivalent inductance $L_{SSPD}=N \cdot L_{kin}$, where $L_{kin}$ is the kinetic inductance of each section of a N-PND of the same area, but for the N-PND $L_{N-PND}=L_{kin}/N$, which reduces the recovery time by a factor $N^2$. For a traditional 10x10 μm$^2$ SSPD the pulse width would be of the order of 10 ns FWHM [19], so the recovery time of the 4-PND shown on the inset of Fig. 2c is indeed a factor ~4$^2$ faster, as predicted.

## 6.2. Proof of PNR capability

In order to infer whether a PND is able to measure the number of incoming photons, it can be probed with a poissonian distribution $S(m)=\mu^m \cdot \exp(-\mu)/m!$ ($\mu$: mean photon number). The limited efficiency $\eta < 1$ of the detector is equivalent to an optical loss, and reduces the mean photon number to: $\mu_e = \eta\mu$. In the regime $\mu_e \ll 1$, $S(m) \sim \mu_e^m/m!$, and for $\mu_e$ low enough eq. (1) can be written as:

$$Q(n) \sim P(n|n) \cdot S(n) \propto \mu_e^n/n! \qquad \text{for } \mu_e \ll 1 \qquad (6)$$

Consequently, the probability $Q(1)$ of detecting one photon is proportional to $\mu$, $Q(2)$ is proportional to $\mu^2$, and so on.

A 5-PND-R was tested with the coherent light from a 850 nm GaAs pulsed laser. The photoresponse from the device was sent to the 150 MHz counter. The photocounts were measured as a function of the threshold level, for different light powers (fig. 5a). As expected, plateaus appear, corresponding to intervals of the output voltage between the well-defined n-photon levels (see fig. 4). In order to measure the count rates for each level, the thresholds of the counter threshold were chosen so that for any $n$ and light intensity the counts corresponding to an $n$-photon absorption event were significantly higher than the counts corresponding to the absorption of more than $n$ photons. In fig. 5b the corresponding detection probabilities relative to one-, two- and three-photon absorption events are plotted for $\mu$ varying from 0.15 to 40.. As $\eta$ is a few percent ($\eta \sim 2\%$) and $\mu$ is a few tens, the condition $\eta\mu = \mu_e \ll 1$ is verified and eq. (6) is therefore valid. Indeed, the fittings clearly show that $Q(1) \propto \mu$, $Q(\mu,2) \propto \mu^2$ and $Q(\mu,3) \propto \mu^3$, which demonstrates the capability of the detector to resolve one, two and three photons simultaneously absorbed.

## 6.3. Conditional probability matrix

The typical application of a PNR detector is the reconstruction of an unknown incoming photon number distribution $S(m)$. It has been shown [22] that $S(m)$ can be recovered given $Q(n)$ and the matrix of the conditional probabilities. Considering equations (2) to (4), it is clear that $P^N$ can be calculated if the vector of the $N$ different quantum efficiencies $\bar{\eta} = [\eta_i]$ is known. $\bar{\eta}$ can be determined fitting the $Q(n)$ measured when probing the device with a light whose $S(m)$ is known.

A 5-PND was tested with the coherent emission from a mode-locked Ti:sapphire laser, whose photon number probability distribution is Poissonian and could be fully characterized by the mean photon number $\mu$ with a power measurement. To determine $Q(n)$, histograms of the photoresponse voltage peak $V_{pk}$ were built for values of $\mu$ ranging from ~1 to ~100. The signal from the device was sent to the 1 GHz oscilloscope, which was triggered by the synchronization generated by the laser unit. The photoresponse was sampled for a gate time of 5ps, making the effect of dark counts negligible. The discrete probability distribution $Q(n)$ was reconstructed from the continuous probability density $q(V_{pk})$ fitting the histograms to the sum of 6 gaussian distributions (corresponding to the five possible pulse levels plus the zero level) and calculating their area (fig. 6). The experimental probability distribution $Q(n)$ measured for different $\mu$ was then fitted to the one predicted by the analytical model (section 3.2) using the vector $\bar{\eta}$ as free parameter (Fig. 7). The value of $\bar{\eta}$ obtained from the fitting became then an input parameter in the Monte Carlo simulations (section 3.3) used to calculate the detected photon number distribution for each value of $\mu$. The three sets of values for the photocount statistics of six levels are in good agreement over almost two orders of magnitude of $\mu$, confirming the validity of the analytical model. Additionally, the fitted efficiencies are rather uniform (2.9±0.5%), indicating a high-quality fabrication process.

## 7. Discussion

The counting capability $M_{max}$ of a PNR detector is limited by several factors. One is the quantum efficiency. From equation (3), assuming the detector saturation is negligible ($n \ll N$) and that all the branches are equal ($\eta_i=\eta$), the probability $Q(n)$ of detecting $n$ photons is proportional to $\eta^n$. In the PNDs tested $\eta$ was a few percent, due to unoptimised film thickness and device design.. Nevertheless, the $\eta$ of SPDs based on the same detection mechanism can be increased up to ~60% [23], and could potentially exceed 90% using optimized optical cavities.. The second limitation is the electrical noise of the detector. As the currents from the branches of the $N$-MPD are summed up to build the output, pulse height discrimination is used to achieve photon number resolution. This makes the noise performance of the device critical for its counting capability as independent noisy signals are summed. Pulse height discrimination can indeed be performed as long as the noise remains lower than the one-photon signal amplitude. In other avalanche-based PNR detectors [2-4, 7-10] the amplitude of the output signal is directly proportional to the number of carriers generated by single photon absorption events through a multiplication process which is intrinsically noisy. The noise on the multiplication gain is then completely transferred to the signal, which is then affected by a fluctuation of the same order. With PNDs, the noisy avalanche carrier-multiplication process [24] causes a fluctuation only in the resistance $R_{hs}$ of the branch driven normal after the absorption of a photon and not in the output current. The amplitude of the photocurrent spike is indeed determined by the partition between the fluctuating resistance $R_{hs}$~1 kΩ and a resistance $R_A$ almost 2 orders of magnitude lower, which is of fixed value. Comparing the broadening of the histogram peaks relative to different numbers of detected photons $n$ (fig. 6), no multiplication noise buildup is observable, as the variance of the peak does not increase with $n$. The broadening of the peaks is then exclusively due to electric noise originating from amplifiers and is not a fundamental property of the detector. To a good approximation the excess noise factor $F$ [25] of the PND is then close to unity and is not limiting $M_{max}$, which is not the case for most of the other approaches to PNR detection [2-5, 7-10]. A third limitation to the counting capability of PNDs is the leakage current $\delta I_{lk}$, which limits the number of parallel wires. However, this issue can be overcome by switching from voltage- to current read-out (e.g. using a transimpedance amplifier), thus decreasing the load impedance.

## 8. Conclusion

In conclusion, a new photon number resolving detector, the Parallel Nanowire Detector, has been presented, which significantly outperforms existing approaches in terms of sensitivity, speed and multiplication noise in the telecommunication wavelength range. In particular, it provides a repetition rate (80 MHz) three orders of magnitude larger than any existing detector at telecom wavelength [2, 5, 11], and a sensitivity [26] one-two orders of magnitude better (with the exception of transition-edge sensors[5], which require a much lower operating temperature). The high repetition rate and high sensitivity make it already suitable – for the first time – for replacing correlation set-ups in quantum optics experiments at telecommunication wavelengths. As compared to SPD [14] arrays, this approach allows a much simpler read-out and is thus scalable to the measurement of photon numbers >2. Indeed, by increasing the efficiency, the performance needed for the single-shot measurement of photon number, as needed in many quantum communication and computing protocols, can be reached. Finally, increasing the maximum photon number to 20-30 photons, the PND could be used as an "analog" detector with single-photon sensitivity, bridging the gap between conventional and single-photon detectors.

## Acknowledgments

This work was supported by the Swiss National Foundation through the "Professeur boursier" and NCCR Quantum Photonics programs, EU FP6 STREP "SINPHONIA" (contract number NMP4-CT-2005-16433), EU FP6 IP "QAP" (contract number 15848), the grant "Non-equilibrium processes after IR photon absorption in thin-film superconducting nanostructures" of Russian Agency on education and Russian Foundation for Basic Research grant 07-02-1362a. The authors thank B. Deveaud-Plédran, B. Dwir and H. Jotterand for useful discussion and technical support and the Interdisciplinary Centre for Electron Microscopy (CIME) for supplying TEM and SEM facilities. A. Gaggero gratefully acknowledges a PhD fellowship at University of Roma TRE.

Figure 1. Scanning electron microscope (SEM) image of a PND with $N$=14 (14-PND, a) and of a PND with $N$=8 and series resistors (8-PND-R, b) fabricated on a 4nm thick NbN film on MgO. The nanowire width is $w$=100 nm, the meander fill factor is $f$=40%. The detector size is 10x10 μm$^2$. The devices are contacted through 70nm thick Au-Ti pads, patterned as a 50 Ω coplanar transmission line. The active nanowires (in color) of the PND-R are connected in series with Au-Pd bias resistors (in yellow). The floating meanders at the four corners of the PND-R pixel correct for the proximity effect.

Figure 2. a, Circuit equivalent of a 4-PND-R. The parameters of the circuit are: $L_{kin}$ =50 nH, $R_{hs}$=1 kΩ, $R_n$=100 kΩ, $R_0$=50 Ω and $R_A$=50 Ω. b, Simulation result for the time evolution during photodetection of the currents $I^{(i)}$ ($i$=1,…,4) flowing through the branches of the device and of the current $I_{out}$ flowing through $R_A$. In the simulation four light pulses (containing one, three, four and two) are sent to the device, with a repetition frequency of 26MHz. The curves relative to $I_{out}$ and the $I^{(i)}$ are plotted in the same units.

Figure 3. Optical equivalent of an $N$-PND. a, The parallel connection of $N$ nanowires is equivalent to a balanced lossless $N$ port beam splitter (BS), every channel terminating with a single photon detector (SPD). b, $k^{th}$ possible configuration of $n$ fired (in black) and $N$-$n$ unfired (in grey) sections. Each incoming photon is equally likely to reach one of the $N$ SPDs (with a probability 1/$N$). Each SPD can detect a photon with a probability $\eta_i$ ($i$=1,..,$N$) different from all the others, and gives the same response for any number ($m \geq 1$) of photons detected.

Figure 4. a, Photoresponse transients taken with a 40 GHz sampling oscilloscope while probing a 10x10 μm$^2$ 4-PND-R in the cryogenic probe station under illumination with 1.3 μm, 100ps-long pulses from a laser diode, at a repetition rate of 26MHz. The solid curves are guides to the eyes. b, Single-shot oscilloscope trace during photodetection by a 8.6x8 μm$^2$ 5-PND. The device was tested under uniform illumination in a cryogenic dipstick dipped in a liquid He bath at 4.2 K. The light pulses at 700 nm form a mode-lock Ti:sapphire laser had a repetition rate of 80 MHz. The five successive response pulses have clearly five discrete amplitudes.

Figure 5. a, Photocounts vs threshold level at different light powers for a 10x10 μm$^2$ 5-PND-R. The device was tested under uniform illumination in a cryogenic dipstick dipped in a liquid He bath at 2.2 K. The light pulses at 0.85μm form the GaAs pulsed laser were 30 ps wide and the repetition rate was 100kHz. The power level was set with a variable fiber-based optical attenuator. The photoresponse from the device was sent to the 150 MHz counter. b, Detection probabilities relative to the one (squares), two (triangles) and three-photon (stars) absorption events as a function of the mean photon number per pulse $\mu$.

Figure 6. Histograms of the photoresponse voltage peak. Histograms were built by sampling the photoresponse of an 8.6x8 μm$^2$ 5-PND. The device was tested under uniform illumination in a cryogenic dipstick dipped in a liquid He bath at 4.2 K. The light pulses at 700nm form a mode-lock Ti:sapphire laser were 40ps wide (after the propagation in the optical fiber) and the repetition rate was 80MHz. The average input photon number per pulse $\mu$ was set with a free space variable optical attenuator. Increasing $\mu$, form 1.5 (a) to 64.9 (l), the shape of the histograms changes as the probability to observe higher response amplitudes increases. The solid lines are the experimental histograms. The dashed lines represent the fitted gaussian distribution of each possible pulse level.

Figure 7. Experimental, fitted and simulated probability distributions. The experimental discrete probability distribution $Q(n)$ (white bins) was estimated from the continuous probability density $q(V_{pk})$ of fig. 6. The 5-PND was probed with several incident mean photon numbers $\mu$: 1.5, 2.8, 4.3, 5.3, 7.7, 12.5, 15.9, 26.9, 33.6, 64.9. The experimental values for $Q(n)$ were then fitted (gray bins) using a genetic algorithm to recover the vector of quantum efficiencies $\bar{\eta}$. The value of $\bar{\eta}$ obtained was used to calculate the detected photon number distribution for each value of μ by Monte Carlo simulations (black bins).

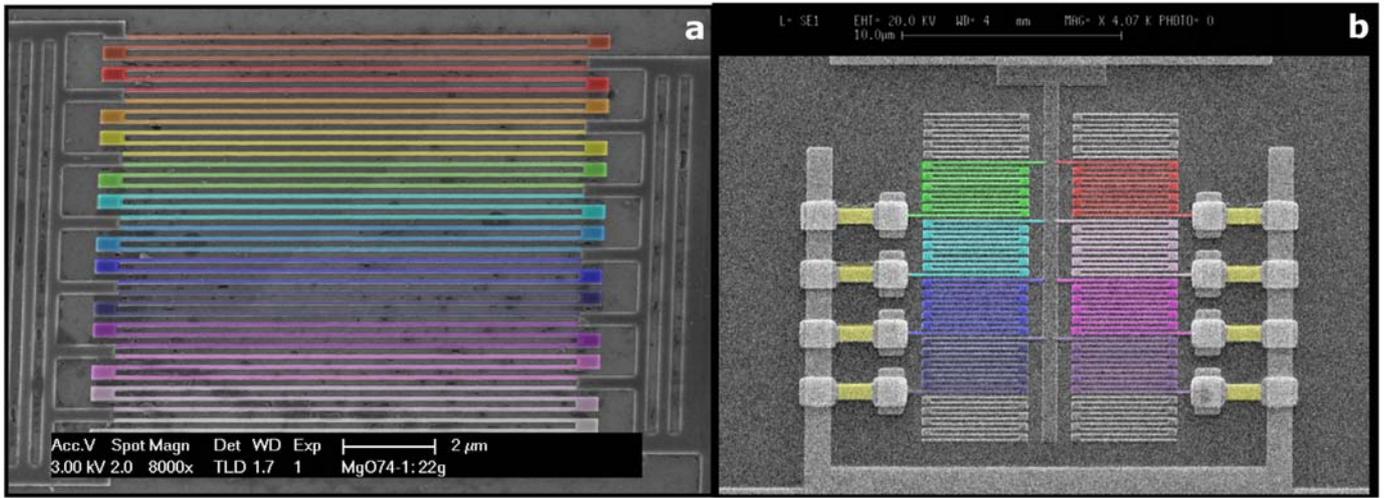

Figure 1

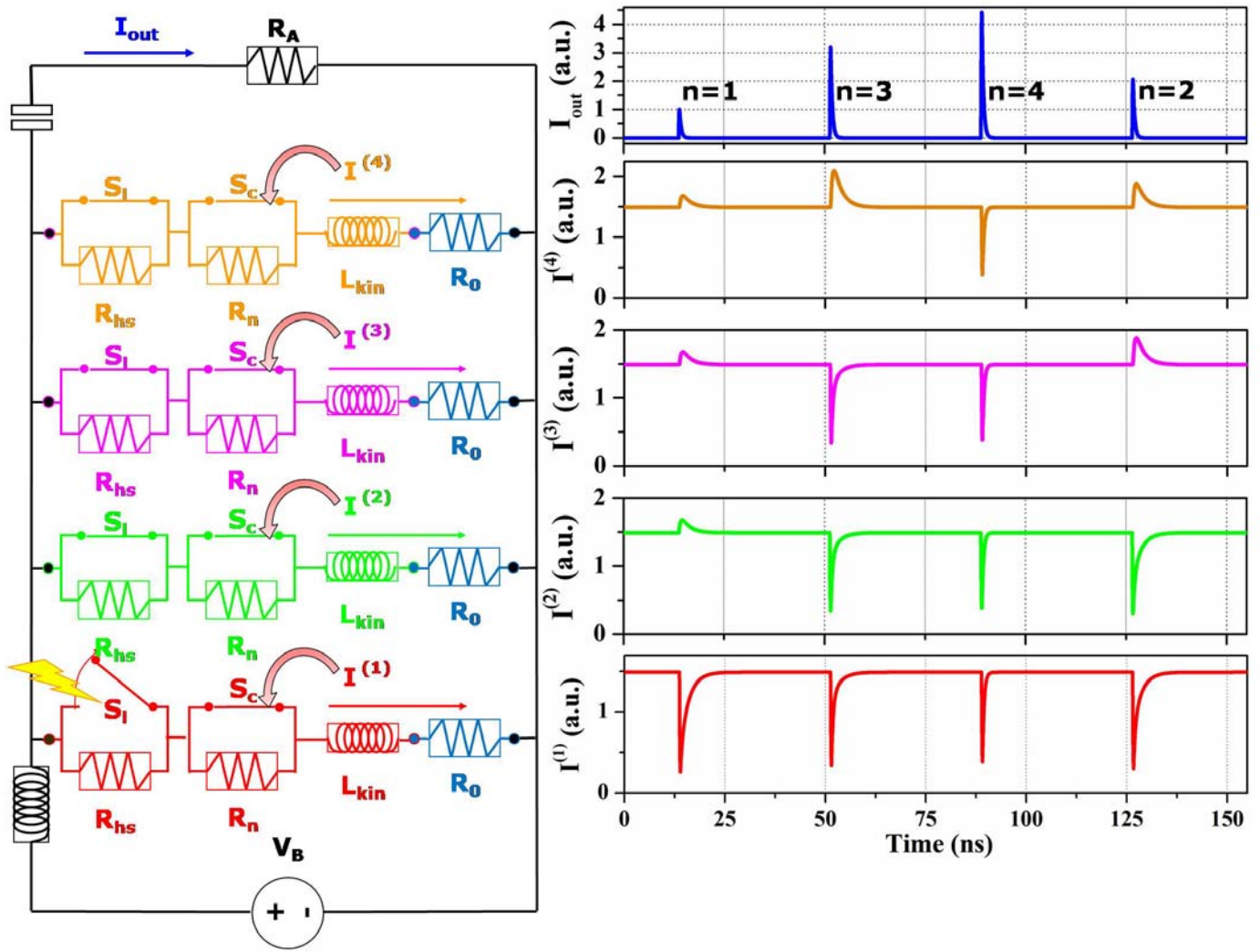

Figure 2

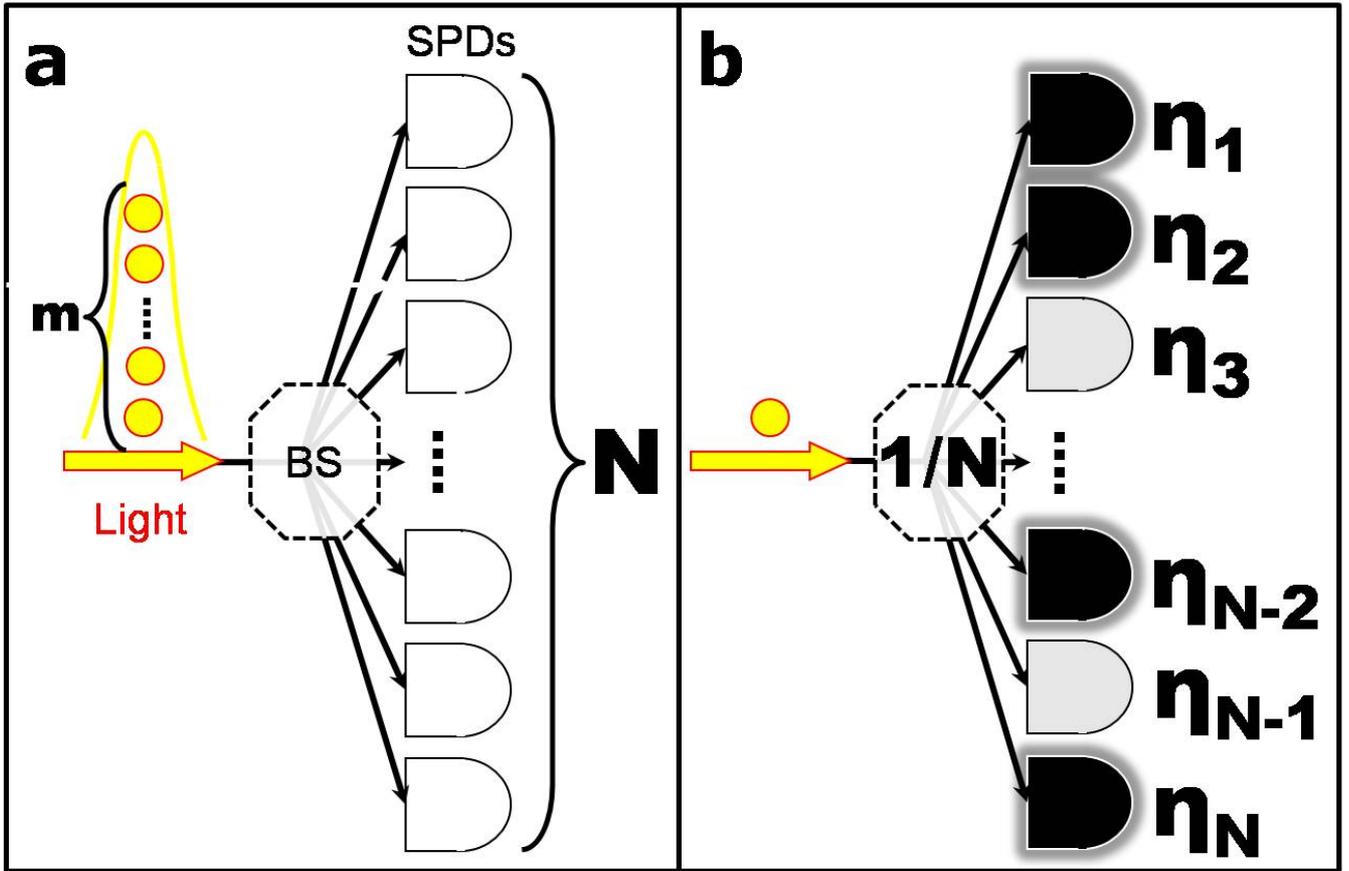

Figure 3

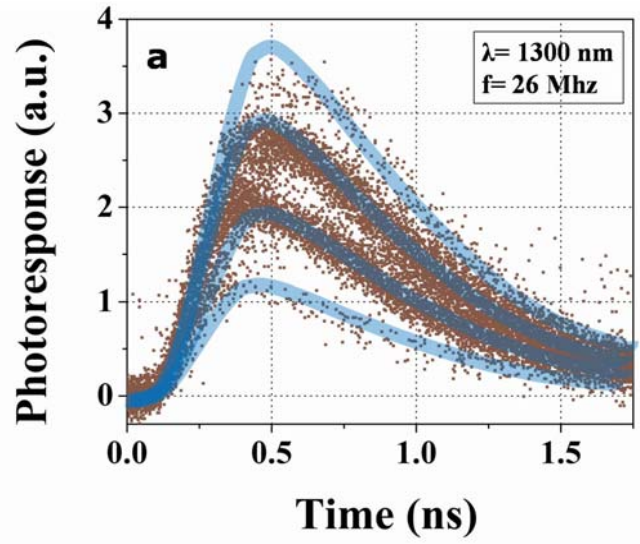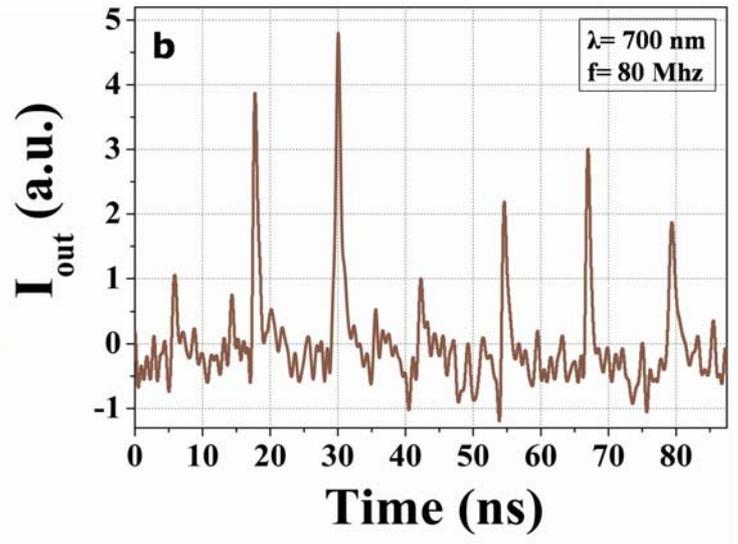

Figure 4

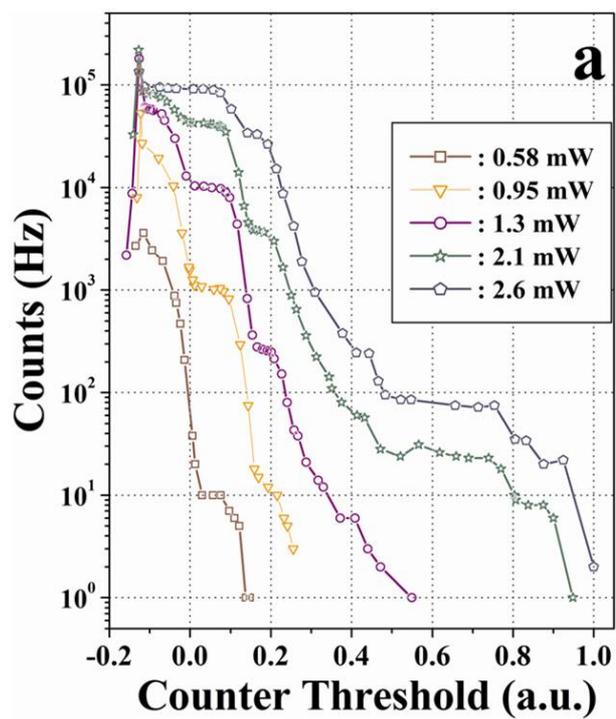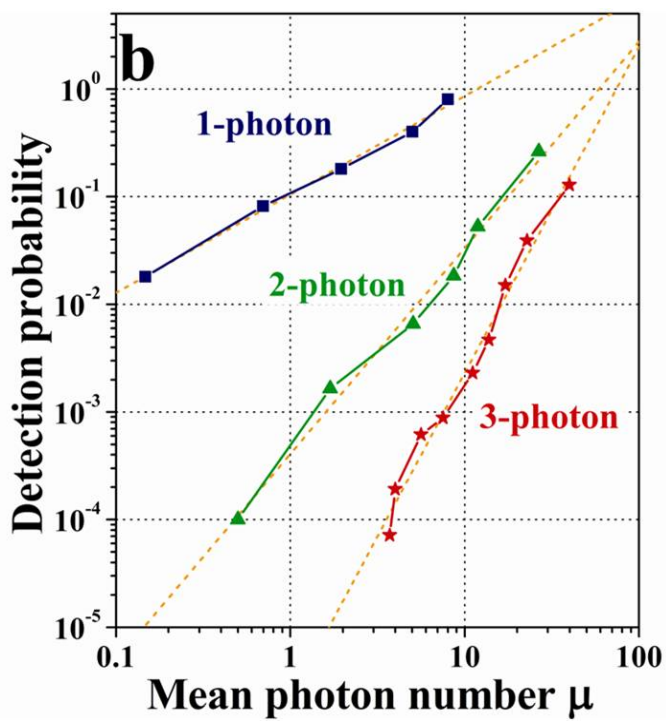

Figure 5

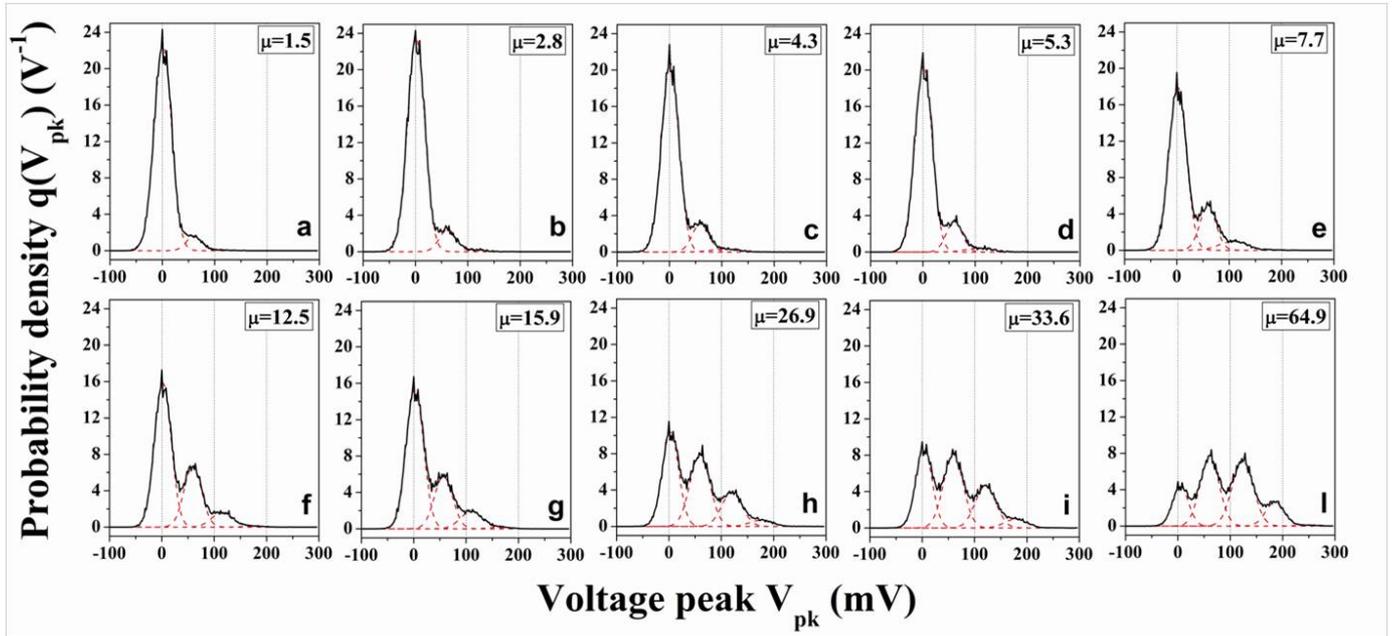

Figure 6

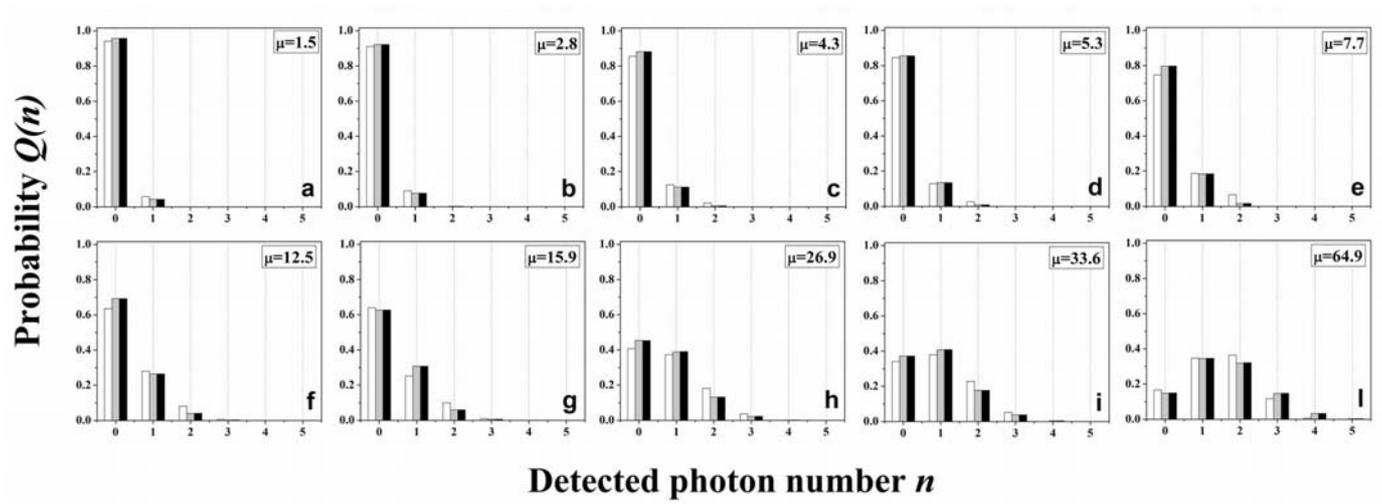

Figure 7